\documentclass[preprint, prX]{revtex4-1} 

\usepackage{hyperref}
\usepackage{amssymb}
\usepackage [dvips]{graphicx}
 \usepackage{graphicx}\usepackage{color}
 \usepackage {amsmath}

\usepackage{dcolumn}%
\usepackage{bm}%

\begin{document}  

  \title{
A study of the fluctuations of the optical properties of a turbid media through Monte Carlo method
}
\author{Emiliano Ter\'an-Bobadilla}
\email{eteran@uas.edu.mx}
\affiliation{Facultad de Ciencias F\'isico-Matematicas Universidad Aut\'onoma de Sinaloa, 80010,
Culiac\'{a}n, Sinaloa, M\'{e}xico.}

\author{Eugenio Rafael M\'endez M\'endez}
\affiliation{Departamento de \'Optica, Centro de Investigaci\'on Cient\'ifica y de Educaci\'on Superior de Ensenada}
\date{\today}

\begin{abstract}

In this work we present a theoretical study on the propagation of light in heterogeneous systems with fluctuating optical properties. To understand the consequences of the fluctuations we perform numerical calculations with uniform and non uniforms systems using Monte Carlo simulations.  We consider two distributions to represent a non-uniform medium: delta function  and an exponential negative distributions.The results show that even with finite moments distributions, may require a large number of interactions for a convergence towards Gaussian statistics. This can be important when estimating the optical properties of thin films.

\end{abstract}
\maketitle


\section{Introduction\label{sec:intro}}

Inhomogeneous medium is characterized by having variations in refractive index or have inclusions of particles with another index. Normally, it is assumed that the media  has properties which do not vary with average position. However, many of these systems, for example, composed of a suspension of particles, tend to form lumps or regions where the particle density is higher (by sedimentation, for example). This breaks down with the assumption of uniformity of the system (see Figure 1) and involves certain difficulties to study. A major consequence is that, depending on the type of fluctuation, random process may not be stationary and hence not ergodic. This means that sample averages aren't equivalent to ensemble averages (ensemble). This class of disordered systems are known as superdifusive media \cite{3,6} and have raised recent interest in different areas of science, in particular in optics \cite{4}.

\begin{figure}[t]
\centering
\includegraphics[scale=.4]{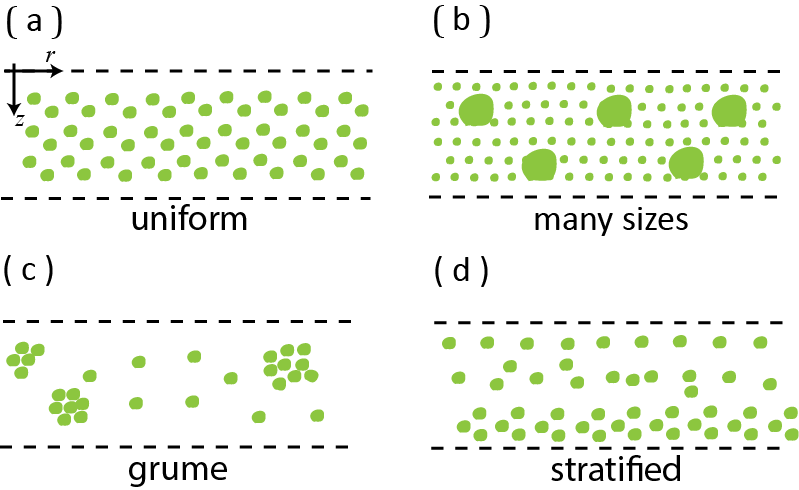}  
\hspace{.1cm}  
  
\caption{Schematic view of a non homogeneous media with a uniform  (a) and non-uniform \mbox{(b)-(d)} particle distribution.
\label{fig:no_uni}
}
\end{figure}

\begin{figure}[t]
\centering
\includegraphics[scale=.45]{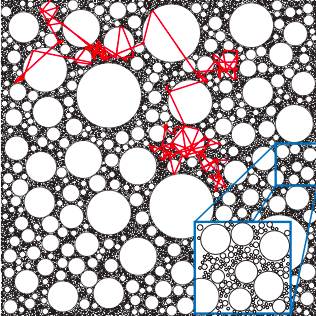}  
\hspace{.1cm}  
  
\caption{Illustration of possible optical paths in a non-homogeneous media with fractal characteristics due to the dispersity in the particle sizes \cite{4}.
\label{fig:fli}
}
\end{figure}

Figure \ref{fig:fli} illustrates the type of paths we can expect in a medium with particles of many sizes. We see that, in regions with small particles, the light follows a zigzag path with step short while to encounter a big particles, abruptaly,  the  paths are so long. This means that in the media there is a non-negligible probability that the light suddenly run, a much longer than the other.

The paper is organized in the following way. In the first section we describe in broad terms the problem we will address. Section II discusses the stochastic properties of a medium where the statistics are Gaussian and non-Gaussian. Section III presents a study of the resultant of flight on uniform and disordered media. Finally, the section provides a summary and conclusions of this work.

\section{Statistical fluctuations of the optical properties}

In simplest theories, which are approximations to the radiative transport equation, inhomogeneous medium can be characterized by their interaction coefficients $\mu_a$ and $\mu_s$,  and also by the anisotropy parameter $g$. These coefficients represent, respectively, the scattering probability per unit length ($\mu_s$), the probability of absorption per unit length ($\mu_a$) and average scattering angle after an interaction ($g = <\cos \theta_s>$). The total interaction coefficient $\mu_t = \mu_s + \mu_a$ and the mean free path is $l = 1/\mu_s$.

In these simple theories it's has, moreover, that  $\mu_s$ and $g$ always appear as $\mu_s' = \mu_s (1 - g)$, which is known as the reduced scattering coefficient. Henceforth, we assume that the medium is not absorbing, so that we can $\mu_a = 0$ and $\mu_t = \mu_s$ and  also it's assume that $g = 0$ (isotropic scattering).

Generally it's assumed that the parameter $\mu_s$ is constant. However, this is not always true in experimental systems. Under the assumption of independent scattering of the particles, the interaction coefficient can be written as
\begin{equation}\label{eq:mut}
\mu_t=\rho C_t,\mbox{    }[\rm{cm}^{-1}]
\end{equation}
where $C_s$ denotes the scattering cross section of particles and $\rho$ indicates the bulk density, \emph{i.e.},  the number of particles per unit volume. We can see that if there are changes in the density, or if the particles have different properties, $\mu_s$ could be a function of position.

We can see from equation (\ref{eq:mut}) that $\mu_t$ depends on the optical size of the particles $C_t$ and of the density $\rho$. We start  the study considering fluctuations with respect  of one the parameter these parameters. Then we will consider the effect of the fluctuations of both of them in the total interaction coefficient. 

\subsection{Uniform systems}

We consider first the case of a non-homogeneous system on a microscopic level but that, beyond a certain level, not changed in their optical properties. We model the media using identical particles with a uniform particle density. The probability of interaction per unit length along a line is then constant.

Let $\mu_s$  the probability of interaction per unit length of a photon in the media. This probability can be written as the sum of the probabilities of scattering and absorption: $\mu_t = \mu_s + \mu_a$. For simplicity, we consider that there is no  absorption in the media so that $\mu_t = \mu_s$. We also define the mean free path between interactions as $l = (\mu_s)^{-1}$.

Let $F(s)$  the probability that a photon, which begins at $s = 0$,  does not scattered on the length $s$. The probability of scattering in a differential length $ds$ is $\mu_s ds$, so that the probability of no scattering at $ds$ is $(1 - \mu_sds)$. We then have
\begin{equation}
F(s+ds)=F(s)(1 - \mu_sds),
\end{equation}
so we can write the differential equation
\begin{equation}
dF(s)=-F(s)\mu_sds.
\end{equation}
The solution, gives the probability that a photon is not scattered in the length $s$,
\begin{equation}
F(s)=\exp(-\mu_ss),
\end{equation}
so that the probability of scattering at this length is given by
\begin{equation}
P(s)=1-\exp(-\mu_ss),
\end{equation}
The probability density function (PDF), $p_s (s) = dP (s) / ds$, which governs the interaction of photons with the media may then be written in the form
\begin{equation}\label{eq_6}
p_s(s)=\mu_s\exp(-\mu_ss)=\dfrac{1}{l}\exp(-s/l),
\end{equation}
The moments of the distribution are given by [7]
\begin{equation}\label{eq_7}
\langle s^n\rangle=\int s^np(s)ds=n!l^n,
\end{equation}
so that
\begin{equation}\label{eq_8}
\langle s\rangle=l, \mbox{  \hspace{1cm}      }<s^2>=2l^2,\mbox{        }\dots,
\end{equation}
and standard deviation
\begin{equation}\label{eq_9}
\sigma_s=\sqrt{\langle s^2\rangle-\langle s\rangle^2}=l.
\end{equation}

\subsection{Systems with two types of particles}

In a non-uniform system it's present variations of the  interaction parameter. We call this parameter fluctuating $\nu$, which its average is $\mu_s$.

The probability density function of displacement in a given region will depend on the specific value of the random variable $\nu$ takes. From the equation (\ref{eq_6}), then the conditional PDF can be writen as 
\begin{equation}\label{eq_11}
p_s(s | \nu) =\nu  \exp(-\nu s).
\end{equation}
Denoting by $p_{\nu} (\nu)$ the PDF to scattering coefficients, we can write
an expression for the new PDF to displacements,
\begin{equation}\label{eq_12}
p_s(s) =\int p_s(s | \nu)p_{\nu}(\nu)d\nu.
\end{equation}
In the case where the scattering coefficient, $\nu$, take just two random values, we can write the PDF as,
\begin{equation}\label{eq_13}
p_{\nu} (\nu) = a \delta(\nu-\mu_1) + b \delta(\nu- \mu_2),
\end{equation}
where $a$ is the probability of the coefficient $\mu_1$ and $b$ represents the probability of occurring $\mu_2$. It's necessaryy    that $a + b = 1$ and  $\mu_s=a\mu_1 + b\mu_2$.

At this point, it is necessary to mention that the modeled system is not a homogeneous mixing of two components (in which case, $\mu_s$ would be equal to the sum $\mu_1 + \mu_2$), but a system with regions with properties $\mu_1$ and other with  properties $\mu_2$.

It's convenient to define also a parameter $\alpha$, by the relationship
\begin{equation}\label{eq_14}
\mu_1 = \dfrac{\alpha}{a}\mu_s,
\end{equation}
implies that,
\begin{equation}\label{eq_15}
\mu_2 = \dfrac{1-\alpha}{1-a}\mu_s.
\end{equation}
This means that the PDF (\ref{eq_13}) can be specified by the parameters $a$, $\mu_1$ and $\mu_2$ or, alternatively, by $\mu_s$, $a$ and  $\alpha$.

The PDF for the displacements, or flights, can be determined by equations
(\ref{eq_11}), (\ref{eq_12}) and (\ref{eq_13}), and is given by
\begin{eqnarray}\label{eq_16}
p_s(s)&=&\int \nu e^{-\nu s}\left[a\delta (\nu-\mu_1)+b\delta (\nu-\mu_2)\right]d\nu,\nonumber\\
&=&a\mu_1 e^{-\mu_1 s}+b\mu_2 e^{-\mu_2 s}
\end{eqnarray}

We can verify that this FDP is normalized properly and that the first moments are
\begin{equation}\label{eq_17}
\langle s\rangle = \dfrac{a}{\mu_1}+\dfrac{b}{\mu_2},\mbox{\hspace{2cm}}
\langle s^2\rangle = \dfrac{2a}{\mu^2_1}+\dfrac{2b}{\mu^2_2}
\end{equation}

\subsection{Systems with negative exponential fluctuations}

For a medium with a distribution of scattering coefficients negative exponential type, we have that the PDF for $\nu$ can be written as,
\begin{equation}\label{eq_18}
p_{\nu}(\nu)=\beta\exp(-\beta\nu),
\end{equation}
where $\beta = 1/\mu_s$ and $\mu_s$ is the average scattering coefficient of the system.

As in the previous case, the equations (\ref{eq_11}), (\ref{eq_12}) and (\ref{eq_18}) we have that the PDF for flight is given by,
\begin{eqnarray}\label{eq_19}
p_s(s)&=&\int_0^{\infty}\left[ \nu \exp(-\nu s)\right]\left[ \beta \exp(-\beta\nu)\right]d\nu,\nonumber\\
&=&\beta\int_0^{\infty}\nu\exp\left[-(s+\beta)\nu\right]d\nu.
\end{eqnarray}
Evaluating the integral, we have
\begin{equation}\label{eq_20}
p_{s}(s)=\dfrac{\beta}{(s+\beta)^2}=\dfrac{\mu_s}{(1+\mu_ss)^2}.
\end{equation}
\begin{figure}[t]
\centering
\includegraphics[scale=.45]{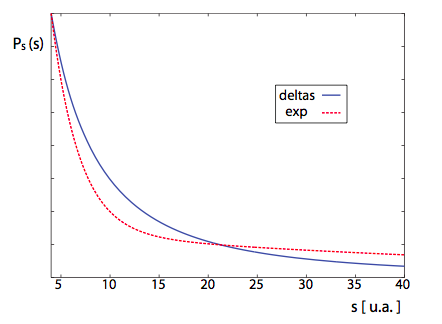}  
\hspace{.1cm}  
  
\caption{Comparing the probability density function defined by equations (\ref{eq_16}) and (\ref{eq_20}). The solid blue curve represents the distribution for flights governed by a variation in $\nu$ double-delta type. The dotted red curve for the negative exponential distribution.
\label{fig_4}
}
\end{figure}

We see that for large s arguments, the probability density behaves
as a Lorentzian L\'evy flight [equation (\ref{eq_10})]. It is worth mentioning that the PDF defined by equation (\ref{eq_20}) has no definite time, which is characteristic of L\'evy flights.

Figure \ref{fig_4} shows the behavior of the density (\ref{eq_16}) and (\ref{eq_20}). The curves were scaled independently to illustrate the differences. We can see that, although the curves appear similar, they have important differences. In particular, it should be noted that the decay of the curve corresponding to negative exponential fluctuations is very slow. 

To better understand the consequences of adopting these PDF, in the next section we present calculations of random walks using Monte Carlo simulations.

 \subsection{Fluctuations in density and optical size}

A realistic system must consider fluctuations in the optical size $C_x$ and density $\rho$.

\section{Random walks\label{sec:medio}}

\begin{figure}[t]
\centering
\includegraphics[scale=.45]{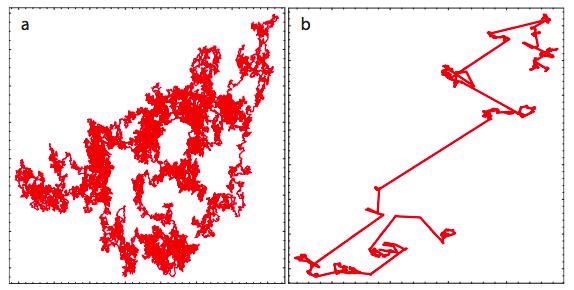}  
\hspace{.1cm}  
  
\caption{Random walk trajectories. (a) Random walk in a uniform media. (b) L\'evy random walk.
\label{fig_5}
}
\end{figure}

In the context of this study, it is interesting to see the result of random walks with different probability density laws described in the previous section, focusing on situations in which the number of interactions is large. The types of situations that may occur in a case with uniform distribution and one in which they occur L\'evy flight is illustrated in Figure 5.

We consider the distance after N number of displacements. For simplicity, we illustrate the method by considering a two-dimensional space and we'll write the total displacement in polar coordinates $(a, \theta)$ (see Figure \ref{fig_5}):
Evaluating the integral, we have
\begin{equation}\label{eq_21}
\mathbf a=ae^{i\theta}=\dfrac{1}{\sqrt{N}}\sum_{k=1}^Ns_ke^{\phi_k}.
\end{equation}
We assume that:
\begin{enumerate}
\item Amplitudes $s_k / \sqrt{N}$ and phases $\phi_k$ are statistically independent.
\item The variables $s_k$ follow the distribution (\ref{eq_6}) with moments given by equation (\ref{eq_7}).
\item The phases $\phi_k$ are uniformly distributed in the interval $(-\pi, \pi)$. This means that the scattering of the particles is isotropic.
\end{enumerate}

We then have that the $x$ and $y$ components are  given by:
\begin{subequations}
\begin{align}
 a_x=&a\cos{\theta}=\dfrac{1}{\sqrt{N}}\sum_{k=1}^Ns_k\cos({\phi_k}),\\
 a_y=&a\sin{\theta}=\dfrac{1}{\sqrt{N}}\sum_{k=1}^Ns_k\sin({\phi_k}),
\end{align}
\end{subequations}
and, with our assumptions, we find that
\begin{subequations}
\begin{align}
 \langle a_x\rangle=0,& \mbox{\hspace{2cm}}\langle a_x^2\rangle=\dfrac{l}{2},\\
 \langle a_y\rangle=0&\hspace{2cm} \langle a_y^2\rangle=\dfrac{l}{2}.
\end{align}
\end{subequations}

\begin{figure}[t]
\centering
\includegraphics[scale=.45]{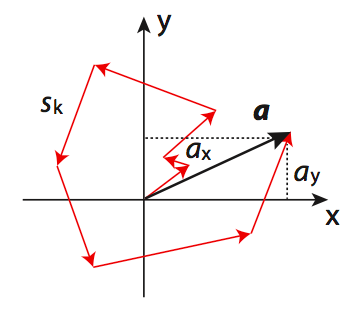}  
\hspace{.1cm}  
  
\caption{Random walk.
\label{fig_5}
}
\end{figure}

When the number of steps, $N$, is very large, the displacement of the photon statistics are Gaussian. That is, both $a_x$ and $a_y$ follow as Gaussian distributions.In this case, the variables have zero mean, second moment $l / 2$ and they are not correlated. It is then circulated Gaussian aletorio process [7]. The joint probability density is then,
Evaluating the integral, we have
\begin{equation}\label{eq_24}
p_{x,y}(a_x,a_y)=\dfrac{1}{\pi l}\exp\left \{-\dfrac{a_x^2-a_y^2}{l}\right\}.
\end{equation}

\begin{table}[b]
\caption{
The system parameters which were performed the  Monte Carlo calculations.
\label{tab:aa}
}
\begin{center}
\begin{tabular}{|l|c|c|c|c|c|}
\hline
layer&$n$ [-]& $\mu_a$ [cm$^{-1}$]& $\mu_s$ [cm$^{-1}$] & $g$ [-]& $d$ [cm]\\
\hline\hline
sup.&1.4&&&&\\\hline
optical system &1.4&0.003& 1867 &0.4&10\\\hline
inf.&1.4&&&&\\
\hline
\end{tabular}
\end{center}
\label{default}
\end{table}%
The length statistics are founded with a probability  transformation to express (\ref{eq_24}) in terms of $(a, \theta)$, and integrating over phase [7], It is found that
\begin{equation}\label{eq_25 }
p_{a}(a)=\dfrac{a}{2 l}\exp\left \{-\dfrac{a^2}{l}\right\},
\end{equation}
for $a> 0$. Then the steps follow a  a Rayleigh distribution. 

We note, however, that our assumptions are valid, moments of the distribution that governs the movement must be finite.

In the next section, we will use Monte Carlo simulations to study three cases, corresponding to the displacement PDF given by equations (\ref{eq_6}),
(\ref{eq_16}) and (\ref{eq_20}), starting with the case of the uniform system.

\begin{table}[b]
\caption{
The system parameters which were performed the  Monte Carlo calculations.
\label{tab:aa}
}
\begin{center}
\begin{tabular}{|l|c|c|}
\hline
\multicolumn{2}{|c|}{Optical properties }&Flights \\
\hline
$\rho$ [cm$^{-3}$]&$C_t$ [cm$^2$]&$p_S(s)$ \\
\hline\hline
cte.&cte.& $\langle\mu_t \rangle\exp\big[-\langle\mu_t \rangle s\big]$ \\\hline
cte.& deltas & $a\mu_1 e^{-\mu_1 s}+b\mu_2 e^{-\mu_2 s}$ \\\hline
cte.&exp. neg. & $\dfrac{\mu_s}{(1+\mu_ss)^2}$ \\\hline
delta&cte.& $\langle\mu_t \rangle\exp(-\langle\mu_t \rangle s)$ \\\hline
delta& deltas & $a\mu_1 e^{-\mu_1 s}+b\mu_2 e^{-\mu_2 s}$ \\\hline
delta&exp. neg. & $\dfrac{\mu_s}{(1+\mu_ss)^2}$ \\\hline
exp. neg.&cte.& $\langle\mu_t \rangle\exp(-\langle\mu_t \rangle s)$ \\\hline
exp. neg.& deltas & $a\mu_1 e^{-\mu_1 s}+b\mu_2 e^{-\mu_2 s}$ \\\hline
exp. neg.&exp. neg. & $\dfrac{\mu_s}{(1+\mu_ss)^2}$ \\\hline
\hline
\end{tabular}
\end{center}
\label{default}
\end{table}%

  \section{MONTE CARLO SIMULATIONs\label{sec:medio}}

To explore the convergence to Gaussian statistics for the three PDF considered, we present calculations based on the MCML (Monte Carlo Multi Layered) simulation [9] using the values for the average properties of the medium shown in Table I. Table defines the parameters of the hypothetical medium  that we study. Assume that the medium is highly scatterer ($\mu_s$ = 1867 cm$^{-1}$) and with low absorption ($\mu_a = 0.003$ cm$^{-1}$) and optically thick $(\mu_sd \gg 1)$, so that a great  number interactions  occur before the photon is lost. To simplify the system, we assumed that the refractive index does not change, we can visualize as if we were immersed in the environment.

To encourage the development of flight we move the point of initial interaction "photons" to the center of the sample and we count $N$ interactions from the origin to obtain the components $a_x$, $a_y$ and $a_z$ of the resultant $\mathbf a$ (see Fig. 6). As discussed earlier, for a large number of interactions, applying the central limit theorem, these components must follow Gaussian statistics.

\subsection{Uniform System}

We study first the uniform system. In this case, the PDF for movement is given by equation (6). Figure \ref{fig_6} shows histograms of the components resulting in, after 10 interactions. The vertical bars represent the histograms of displacement, and red curves, Gaussian functions that adjust data. We can see that although there are slight changes in the heights of the curves all have the same width $\omega_o = 22.5\ $ m. Clearly the components of the displacements resulting in good agreement with the expected Gaussian distribution, which is consistent with the central limit theorem. This, despite the fact that we considered only 10 interactions. 

Figure \ref{fig_7} shows the histogram of the magnitude, $ a$, of the resultant. As expected, the result fits very well to a Rayleigh PDF.

\subsection{System with two types of particles}

\begin{figure}[t]
\centering
\includegraphics[scale=.5]{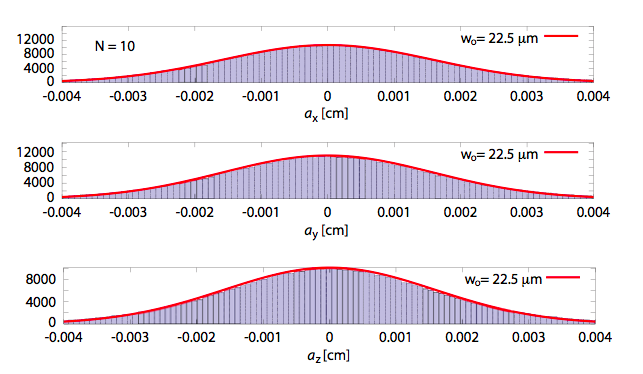}

\caption{
Histogram of the components $a_x$, $a_y$, and $a_z$ of the resultant of the  random walks for  a uniform system.
\label{fig_6}
}
\end{figure}

\begin{figure}[b]
\centering
\includegraphics[scale=.5]{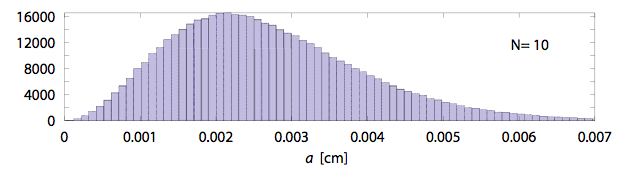} 

\caption{
Histogram of the magnitude of the resultant $a$ of the random walk.
\label{fig_7}
}
\end{figure}
Consider now the case of the medium with two types of particles. FDP for displacement is given by equation (16). Figure \ref{fig_6} shows histograms of the resultant components after 10 interactions. As in previous cases, vertical bars represent histograms and red curves  Gaussian functions adjusted in height. This figure was generated by taking $a = 0.1$ and $\alpha = 0.01$, so that $\mu_1 = 0.1\mu_s$ (long steps with a low probability) and $\mu_2 = 1.1\mu_s$ (short step with high probability). As the difference between the values of  the coefficients of scattering is great, it has the possibility of abrupt fluctuations.

\begin{figure}[t]
\centering
\includegraphics[scale=.5]{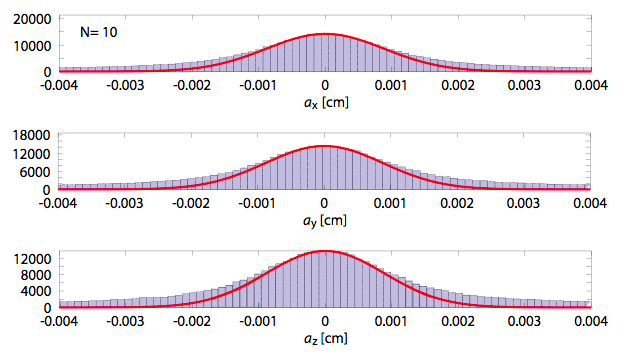}  
\caption{
Histogram of the components $a_x$, $a_y$, and $a_z$ of the resultant of the random walk, considering the PDF given by equation (\ref{eq_16}) with $a = 0.1$ and $\alpha = 0.01$.
\label{fig_8}
}
\end{figure}

Unlike the previous case, Figure  \ref{fig_8} shows that it does not have a good fit to Gaussian curves. This means that after 10 steps, the statistics do not converge to such statistics. Should be noted that, if we increase the number of interactions or flights eventually expected convergence is obtained based on the central limit theorem. On the other hand, keeping the number of flights 10, but considering larger values of $\alpha$ (which implies that the two deltas distribution of  the scattering coefficients are closer) is also obtained convergence Gaussian statistics.

\subsection{System with a  negative exponential distribution of particles}
\begin{figure}[t]
\centering
\includegraphics[scale=.5]{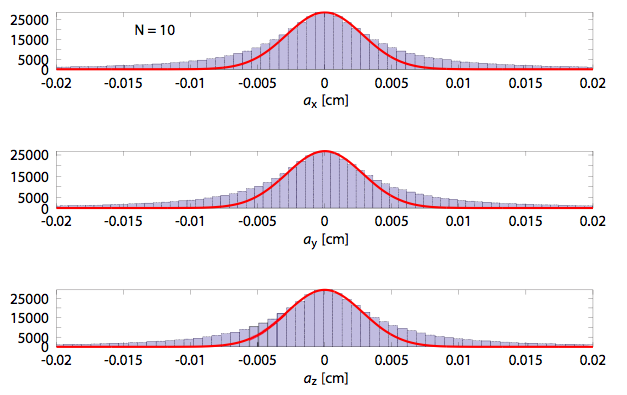}  
\caption{
Histogram of the components $a_x$, $a_y$, and $a_z$ of the resultant of the random walk, considering the PDF given by equation (\ref{eq_20}).
\label{fig_9}
}
\end{figure}

Now we assume that PDF governing displacement is given by equation (\ref{eq_20}). Figure 10 shows the histogram of the components of the resultant, $\mathbf a$, after 10 flights. As in previous figures, vertical bars represent histograms and red curves Gaussian functions adjusted in height .We see that the histograms of the components do not fit Gaussian curves.We can also see that the range of values ??taken by these components is much broader than in the previous cases, indicating that the fluctuations are much larger in flight, and can be up to an order of magnitude larger.

This is not surprising, then flight statistics given by equation (\ref{eq_20}) represent a statistical approach to type L\'evy, and L\'evy type processes are caracterizazdos by violent fluctuations that make the resulting not converge to Gaussian statistics.

\begin{figure}[t]
\centering
\includegraphics[scale=.5]{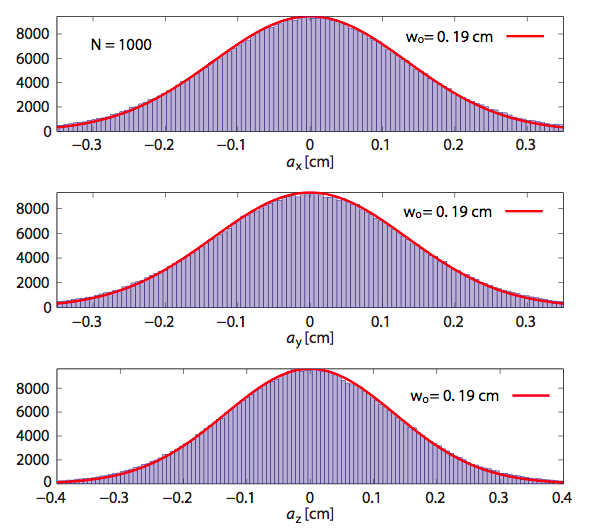}  
\caption{Histogram of the components $a_x$, $a_y$, and $a_z$ the resultant of random walks for $1,000$ flights.
\label{fig_10}
}
\end{figure}

We studied, however, the possibility of convergence after a very large number of interactions. Figure \ref{fig_10} shows the histogram of the resultant components after $1,000$ flights. We see that, after such a large number of flights, the statistics converge if Gaussian statistics appear, although we should mention that this does not necessarily mean that the central limit theorem is valid in this type of situation. The statistics themselves seem to converge to Gaussian statistics, although we should mention that this does not necessarily mean that the central limit theorem is valid in this type of situation.

The above results show that for this medium, if they occur a sufficiently large number of interactions, it will appear to be Gaussian statistics. However, such large fluctuations have important implications finite system, as in films, in which the number of interactions is limited by the film thickness.

\section{COMMENTS AND CONCLUSIONS}

We have seen that the FDP adopted  for the flights can determine the convergence or lack of it to Gaussian statistics.

The fact of having non-Gaussian statistics and the ability to take big steps involve major changes in the properties of a film of this type of media. In these superdifusivos media, for example, the opacity of a film can be lowered considerably. 

The results show that in uniform media, after 10 steps, has a good convergence to Gaussian statistics. In the other two types of media considered more interactions are required to have these statistics. In particular, for the medium with negative exponential fluctuations are required in the order of 1,000 interactions to approach these statistics.

\section{Acknowledgements}

E. T. is grateful to the authorities of the UAS and CICESE for their support to perform this study.  This work has been supported by PROMEP  under grant 2012.

\end{document}